\documentclass{cimento}
\usepackage{graphicx}
\title{{New Methods for Investigating Superconductivity at Very High Pressures}}
\author{Viktor V. Struzhkin\from{ins:x}\ETC ,
Yuri A. Timofeev\from{ins:y},
Eugene Gregoryanz\from{ins:x},
Russell J. Hemley\from{ins:x},
Ho-kwang Mao\from{ins:x}
}
\instlist{\inst{ins:x} Geophysical Laboratory and Center for
High Pressure Research,Carnegie Institution of Washington,
5251 Broad Branch Rd, N. W., Washington D.C. 20015
  \inst{ins:y} Institute for High Pressure Physics, Russian Academy
of Sciences, Troitsk 142 092, Russia }
\begin{document}

\maketitle




\section{Introduction}

As a result of rapid developments in diamond-cell techniques, a
broad range of studies of  the physical and chemical properties of
solids can be now conducted {\it in situ} to  megabar  pressures  
(i.e., $>$ 100 GPa). The highest  superconducting T$_c$ has been 
achieved by application of pressure of $\sim$ 30 GPa on 
HgBa$_2$Ca$_2$Cu$_3$O$_{8+\delta}$ giving critical
superconducting temperature of 164 K \cite{htcp}. The same
material has a record T$_c$=133 K at ambient pressure. It took
almost 30 years after the theory of superconductivity by Bardeen, 
Cooper and Schrieffer (BCS) in 1957
\cite{BCS}, and more than 75 years after discovery of the first
superconductor by Kamerlingh Onnes in 1911 \cite{Onnes} (Hg, with
a T$_c$=4.15 K) to produce the first material that was
superconducting above the boiling point of nitrogen (77.3 K):
YBa$_2$Cu$_3$O$_{7-x}$ with T$_c$=91(2) K \cite{ybco}. Notably, it
was a consideration of the effect of pressure on the 40 K La-cuprates
\cite{presscupr} that guided experiments leading to the 90 K 
YBa$_2$Cu$_3$O$_{7-x}$ superconductor \cite{ybco}.
Such applications of pressure allow to tune over a broad range 
electronic, magnetic, structural and vibrational properties 
of solids \cite{HemAsh,Hemley97}. Earlier techniques have been adapted to
very small sample volumes (typical sample size in high pressure
experiment is about 100x100x20 $\mu$m$^3$ at 30 GPa) to measure
superconducting transitions inductively (by using magnetic
susceptibility methods) or by direct conductivity methods at high pressures. 
The samples must be further reduced in size when a  pressure medium 
is used (which is mandatory for many applications), and  also for 
very high pressure experiments to above 100 GPa. To handle these 
demanding tasks new methods have been developed to measure
superconducting transitions by utilizing Meissner effect in  very
small samples. We will present here in-depth discussion of a
relatively new double-frequency modulation technique. The
technique was used in measurements of T$_c$ in sulfur to
230~GPa without pressure medium \cite{Gregoryanz01}, and in recent 
measurements of superconductivity in MgB$_2$ to 44~GPa in 
He pressure media \cite{strumgb2}.
The direct conductivity methods have recently been extended to pressures 
of 250 GPa \cite{eremets}. 

The chapter is organized as follows. We start with an overview 
of magnetic techniques in Section 2. We then give a detailed description 
of double-frequency modulation technique. In Section 4 we discuss 
the pressure effects on superconductivity in simple s-p metals. 
The T$_c$(P) in the  chalcogens is given in Section 5. In Section 6 
we present  pressure data for the recently discovered 
high-temperature superconductor MgB$_2$.

\section{Overview of existing techniques}

One of the very first techniques for measuring $T_c$(P) 
in superconductors is well described in a review by Klotz, Schilling, 
and M\"uller \cite{klotz1}. This is a single-frequency standard 
technique often used at ambient pressure and adapted for diamond 
anvil cell experiment. 
The technique is capable of detecting signals from the 
samples as small as 80 $\mu$m in diameter. Similar technique was 
developed by Tissen and used in measurements of $T_c$ in La to 
50 GPa \cite{Tissen}. 

Several other techniques have been 
developed in recent years  \cite{Timofeev,Raphael98,Ishizuka95} 
to overcome the problem of the background signal 
in single-frequency technique.  The most notable techniques are
the third-harmonic method \cite{Raphael98}, and 
the vibrating magnetometer technique \cite{Ishizuka95}. 
Both techniques are capable of measuring  signal from very small samples; 
however, the smallest 
possible sample size was not estimated in the original 
publications\cite{Ishizuka95,Raphael98}. The vibrating 
magnetometer technique has been used to detect superconductivity in vanadium 
to 120 GPa where $T_c$=17.2 K \cite{Ishizuka00}.

The method presented here was introduced by Timofeev \cite{Timofeev}. 
We have addressed the techniques for measurements of superconducting 
critical temperatures and Curie temperatures in ferromagnets recently 
\cite{Timofeev99,Timofeev00}. The technique has been used to the highest 
pressure of 230 GPa \cite{Gregoryanz01}, and keeps the record for 
high-pressure magnetic susceptibility measurements of superconducting $T_c$.

The highest reported pressure of 250 GPa for superconductivity measurements 
was reached recently in experiment on boron \cite{eremets} 
by electrical conductivity technique. We refer the reader to original 
papers and reviews on the progress in investigation of transitions from 
insulating to metallic states and $T_c$ measurements in high-pressure 
metals by resistivity methods \cite{eremets,amaya,hemley}.    
Here we will focus on measurements of superconducting transitions by magnetic
susceptibility techniques in diamond anvil cell.

\section{Double-frequency modulation method}

\subsection{Overview of the technique}

\begin{figure}
\includegraphics[width=5in]{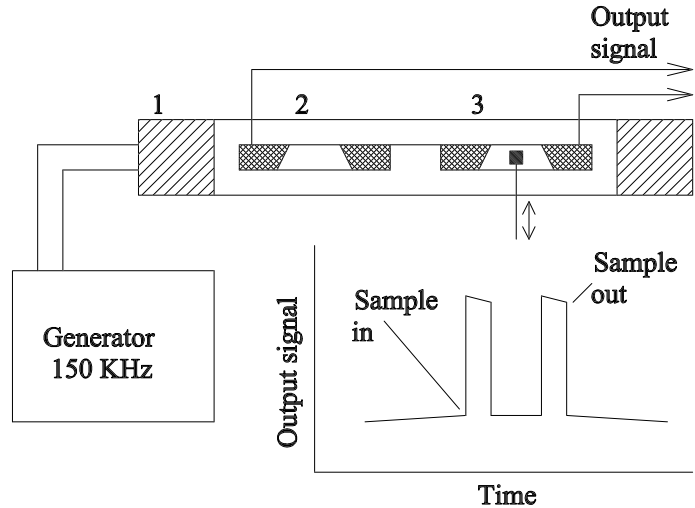} \caption{Schematic representation of
the background subtraction principle in magnetic susceptibility
measurements: 1 - primary coil; 2 - secondary compensating coil; 3
- secondary signal coil. Removal of the sample from the signal
coil produces measurable changes in the total output signal. }
\label{fig1}
\end{figure}

\begin{figure}
\includegraphics[width=3in]{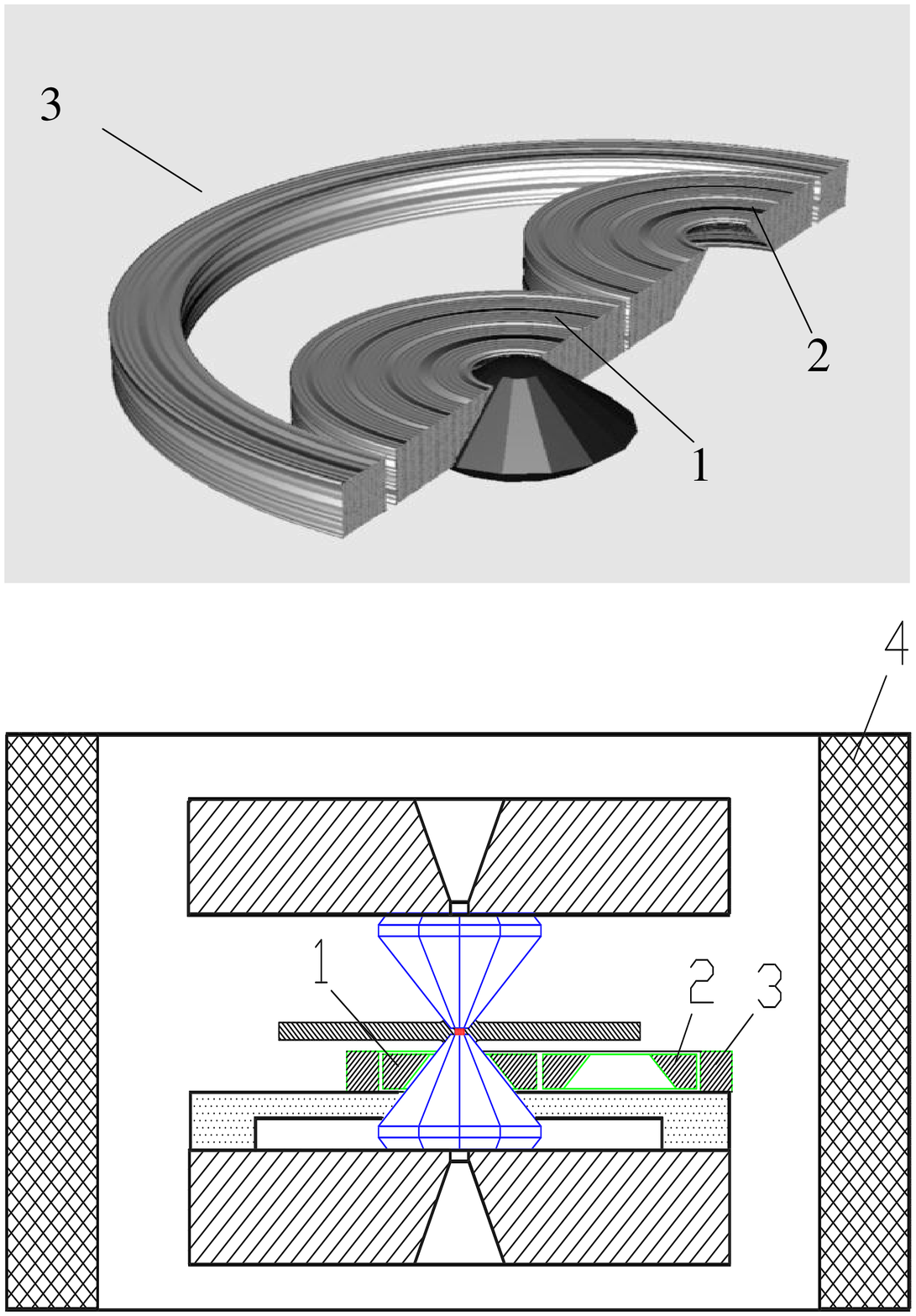}\includegraphics[width=3in]{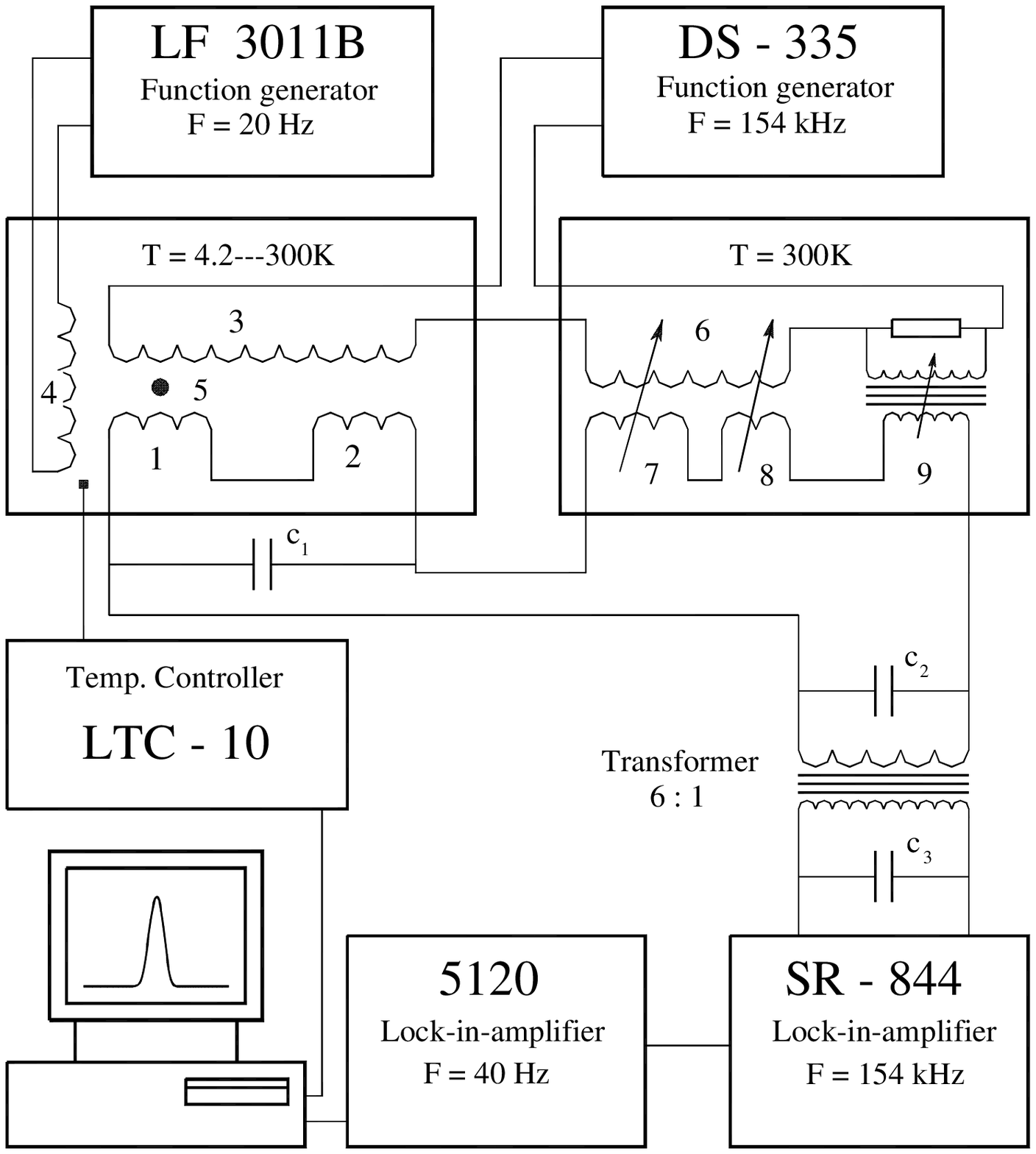}
\caption {Schematic of the double-frequency modulation setup. Coil 4 is
used to apply low frequency magnetic field to modulate the
amplitude response from the high-frequency pick-up coil 2 due to
the superconducting sample. The setup
includes two signal generators and two lock-in amplifiers, 
operating at low (20-40Hz) and high (155KHz) frequencies. Details 
are presented elsewhere \cite{Timofeev00}.
} \label{fig2}
\end{figure}

\begin{figure}
\includegraphics[width=5in]{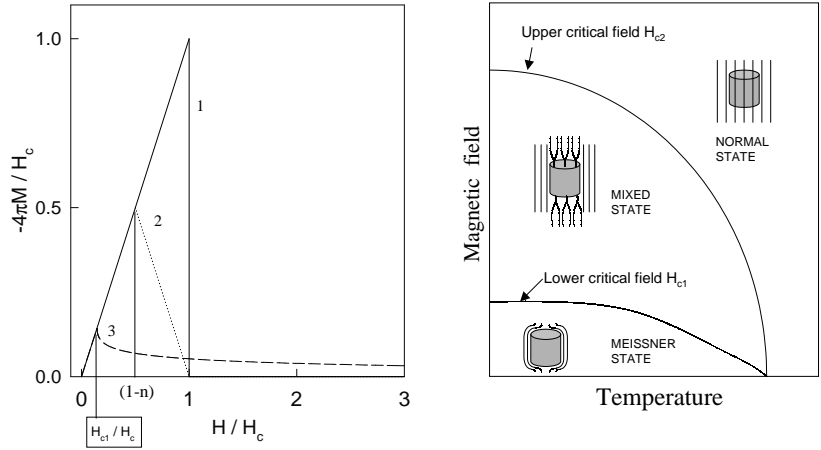}
\caption{Magnetization curves for type I and type II
superconductors. 1 - ideal type I superconductor; 2 - type I
superconductor, with demagnetization factor taken into account; 3
- magnetization curve for type II superconductor. Right panel
shows schematically where Meissner state, mixed state,  and
normal state are located in H-T space for type II superconductor.
It shows also that H$_{c1}$ is almost linear close to T$_c$.}
\label{fig3}
\end{figure}

\begin{figure}
\includegraphics[width=5.5in]{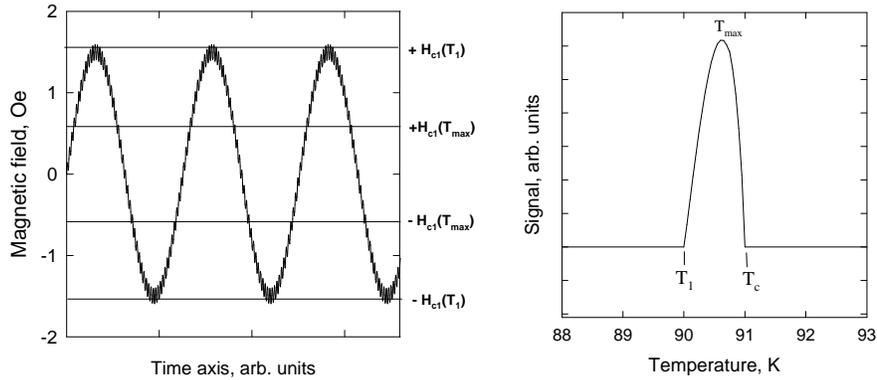} \caption{Left panel: schematic
presentation of magnetic field variation with time near the
sample. The low frequency component at frequency $f$ is used to
``virtually'' remove the sample from the cell: when the amplitude of
the low-frequency magnetic field H$_0$ exceeds the first critical
field H$_{c1}$, the magnetic field penetrates the sample volume.
Right panel: signal at frequency $2f$ extracted from the amplitude
of the high-frequency signal. At $T>T_c$ the signal disappears because
the magnetic field penetrates  the sample volume at all times;
thus there is no variation of high-frequency signal amplitude
with time. When $T< T_1$, signal also vanishes for an ideal
superconductor ($T_1$ is determined by the condition
$H_{c1}(T_1)=H_0$), because then $H_{c1}>H_0$ and magnetic field
is excluded at all times from the sample volume, and there is no
variation of the of high-frequency signal amplitude with time.
When $T_1<T< T_c$, the  magnetic field enters the sample volume
twice per period of the low-frequency magnetic field $T$
(frequency $f=1/T$), thereby producing amplitude modulation of
the high-frequency signal with period $T/2$  (at frequency $2f$).}
\label{fig4}
\end{figure}

\begin{figure}
\includegraphics[width=5in]{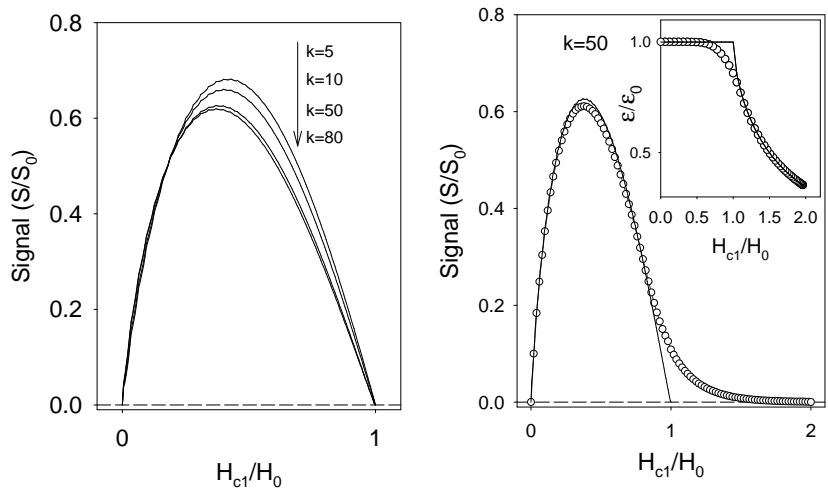} 
\caption{Signal dependence for type II superconductors 
calculated for different values of the Ginzburg-Landau parameter  $k$,
using Hao-Clemm theory (see text). Right panel
illustrates how the signal shape is modified if the sharp cusp in
the magnetization curve at H$_{c1}$ is smeared out  by experimental
conditions (e.g., when the amplitude of the high-frequency
magnetic field is not negligible compared to the amplitude of the
low-frequency field) or by irreversibility effects in
magnetization.} \label{fig5}
\end{figure}

\begin{figure}
\includegraphics[width=5in]{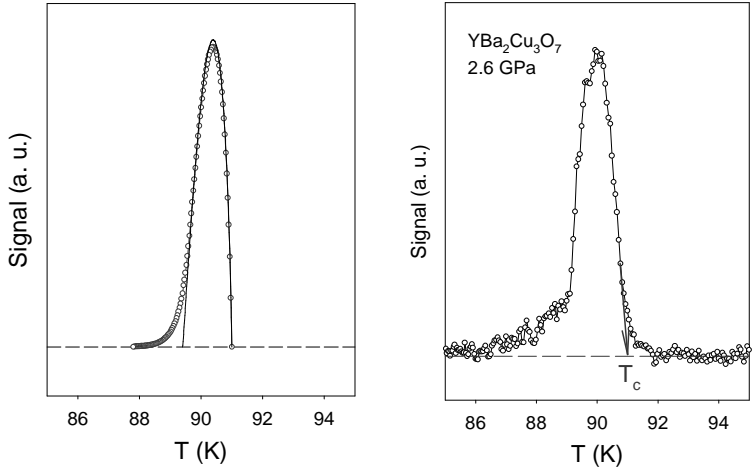} \caption{Comparison of the theoretical signal shape
from Fig. \ref{fig5} (right panel) with experimentally observed
signal shape for slightly overdoped YBa$_2$Cu$_3$O$_{7-x}$ in a He
pressure medium.} \label{fig6}
\end{figure}

\begin{figure}
\includegraphics[width=5in]{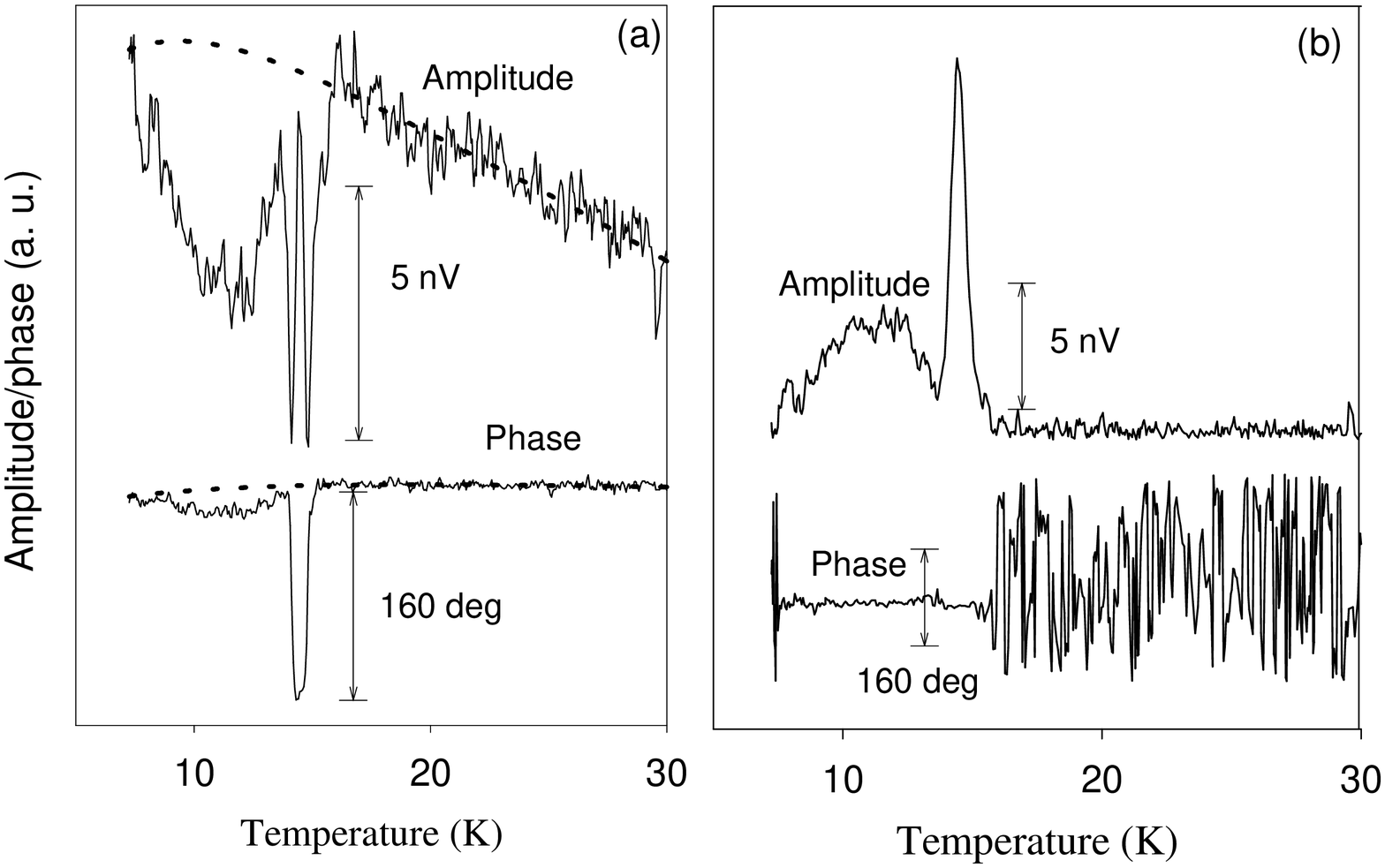} \caption{Superconducting signal
from sulfur sample at 231 GPa, overlapped with the background.
This figure illustrates how smooth background can be subtracted
from the signal, using information for both signal amplitude and
phase. See text for details. } \label{fig9}
\end{figure}

We begin with a
discussion of problems that arise when measuring magnetic
properties of a small superconducting sample  in a  system
consisting of  signal and  compensating (secondary) and exciting
(primary)  inductance coils located in the vicinity of the  sample
(Fig.\ref{fig1}). The exciting coil (1) creates an alternating
magnetic field which produces electromotive forces in both the
signal (3) and compensating (2) coils. These coils are included in
the electrical circuit  in such a manner that their electromotive
forces act in opposite directions and nearly compensate for each
other. The difference between the two electromotive forces
determines the background signal with magnitude depending on
several  factors. The most significant are (a) differences in the 
geometric parameters of the signal and compensating coils and 
their unavoidable asymmetric disposition inside the exciting coil, and
(b) the proximity of the system of coils to metal parts of the
high-pressure chamber (not shown in the figure), which distort the
uniform distribution of magnetic flux passing through  signal and
compensating coils due to geometric asymmetry and electrical
conductivity.

  A modulation technique can be applied to detect the
superconducting critical temperature due to the fact that one can
``virtually'' remove the  superconducting sample from the high
pressure cell by applying an external magnetic field that destroys
superconductivity in the sample.  The technique is based on the
fact that the magnetic susceptibility of superconducting materials
depends on the external magnetic field enclosed in the volume of
the sample. When the magnetic field is high enough to quench the
superconductivity, the Meissner effect is suppressed and the
magnetic field penetrates the sample volume. (This happens at
the  critical magnetic field H$_{c}$). In contrast, the
susceptibility of the metallic parts of the high-pressure cell
(diamagnetic and paramagnetic) is essentially independent of the
external field.  Thus, insertion of the high-pressure cell
containing the sample in an external magnetic field exceeding the
critical value  changes the part of the signal produced by the
sample, while the background remains practically constant.  This
fact allows the separation of the signal arising from the sample
from that of the background.  We apply the low-frequency (f = 22
Hz) magnetic field with an amplitude up to several dozen Oersted,
which causes the destruction of the superconducting state near the
superconducting transition. This in turn leads to a change in the
magnetic susceptibility of the sample from -1 up to 0 twice in a
given period, and produces a modulation of the signal amplitude in
the receiving coils at a frequency 2f.  The subsequently amplified
signal from the lock-in amplifier is then recorded as a function
of temperature on the computer.  The critical superconducting
temperature T$_c$  is then identified as the point were the signal
goes to zero due to the disappearance of the diamagnetic signal at
T$_c$. Fig. \ref{fig2} shows the outline of the system of
coils and relevant electronics for detecting the double-frequency
2f modulated signal.

\subsection{Signal shape}

The dependence of the  signal on temperature for the outlined
double-frequency modulation technique can be estimated using
the Hao-Clemm model for reversible magnetization in type II
superconductor \cite{haoclemm}. Hao-Clemm theory has been highly
successful in describing the magnetization curves of high-Tc
superconductors, and is valid in the temperature and
field region where magnetization is thermodynamically reversible
and fluctuation effects are not important.   Figure \ref{fig3} 
shows the magnetization curves for type I and type II
superconductors. We illustrate the time dependence of
the magnetic field at the sample position in Fig. \ref{fig4}. When
the temperature increases and approaches T$_c$, the critical
magnetic field H$_{c1}$ decreases (almost linearly close to
T$_c$), reaching zero at T$_c$. Using magnetization curves
for type II superconductor we have calculated numerically the
signal at the second harmonic of the low-frequency magnetic field,
which is shown in Fig. \ref{fig5}. The signal starts developing
when H$_{c1}$ is equal to the amplitude of the low-frequency
magnetic field (temperature T$_1$, Fig. \ref{fig4}) and reaches
its maximum at temperature T$_{max}$, which is determined by the
actual values of the parameters involved. The signal drops to
zero at T$_c$. We assumed in the calculation that H$_{c1}$ is a 
linear function of (T$_c$-T) close to T$_c$. 
The numerically calculated signal shape 
does not depend significantly on the value of the Ginzburg-Landau parameter 
and is shown in Fig. \ref{fig5}.  The signal shapes calculated for 
different magnetization curves are close to the signal
shape observed in experiments on YBa$_2$Cu$_3$O$_{7-x}$ in He
pressure medium (Fig. \ref{fig6}).
Our assumptions of reversible magnetization are supported by
reported magnetization measurements in YBa$_2$Cu$_3$O$_{7-x}$
between 80 and 90~K \cite{Xu}. However, more realistic
calculations should take into account the irreversibility effect
on the magnetization curves near T$_c$, similar to calculations
performed in Ref. \cite{shatz} for the signal shape at the third
harmonic of the excitation frequency.

\subsection{Sensitivity}

The crucial parameter for such a system is its sensitivity. We
give below a simple estimate of the sensitivity based on classical
electrodynamics. The sample is very small and acts as a magnetic
dipole  with a magnetic moment (in SI units) $M=V_{s}\chi B/\mu_0$
(where $V_s$ is the sample volume, $\chi$ is  magnetic
susceptibility, $B$ is magnetic induction, produced by excitation
coils, and  $\mu_0$ =4$\pi$ $\cdot$ 10$^{-7}$ H/m is
 the magnetic permeability of a vacuum).
The EMF induced by the sample on the detection coils is

\begin{equation}
\varepsilon=2\pi f{ n \over D} V_s \chi B,
\label{eq:sen}
\end{equation}

\noindent where $f$ is  the frequency of the exciting field,
 $n$ is  number of turns, and $D$ is a diameter of detection coils.
Using experimental parameters
$f$=60 KHz,
$n$=300,
$D$=3.5 mm,
$V_s$=10$^{-14}$  m$^3$ (sample size 33x33x10 $\mu$m$^3$ ),
$B$=3 $\cdot$  10$^{-5}$ T, and assuming that  $\chi$=-1 for $T<T_c$,
we obtain  $\varepsilon$=12 nV.
This estimate is fairly close to experimental values.
Eq.\ref{eq:sen} is also helpful in estimating ways of increasing
 the sensitivity of the experimental setup. It
is evident  that when  parameters  $n$, $D$, $B$ are optimized,
one can increase the exciting frequency $f$  in order to increase
the sensitivity. However, increasing the operating frequency
introduces problems related to the fact that stray capacitances of
all electrical leads and cables become important. The details of
the setup are described elsewhere \cite{Timofeev00}. In Fig.
\ref{fig2} the schematic representation of the  setup is given; 
we refer the reader to Refs.
\cite{Timofeev,Timofeev99,Timofeev00} for details. We will give
below representative examples of T$_c$ measurements at megabar pressures.

\subsection{Background issues}

Samples of 99.9995\% purity S were loaded in Mao-Bell cells
\cite{mao_cell} made from Be-Cu and modified for measurements down
to liquid helium temperatures. The gaskets made from nonmagnetic
Ni-Cr alloy were used together with tungsten inserts to confine
the sample;  no pressure transmitting media were used. The
gasket and insert may be responsible for the temperature dependent
background seen in the raw temperature scans (e.g., Fig.
\ref{fig9}a). Two peaks are clearly seen at $\sim$10-12 K and
$\sim$17~K. The second broad peak at lower temperatures arises
from the sample outside of the flat culet, where pressure is
considerably lower than within the center of the culet.  The
splitting of the  17~K peak is artificial and only reflects the 
fact that the signal
amplitude has increased substantially with respect to the
background.

The background signal in our measurements appeared to be ferro- or
paramagnetic, as its phase is approximately opposite to that of
the signal from the sample (diamagnetic) (see Fig. \ref{fig9}a).
Because the background signal changes smoothly with temperature,
we can separate the signal from the background by the simple
procedure illustrated in Fig. \ref{fig9}. We measure amplitude and
phase of a sum of signal and background with a lock-in
technique. The signal changes very abruptly in the vicinity of the
superconducting transition, allowing us to see these changes both
in amplitude and phase of the signal (Fig. \ref{fig9}a). 
It is straightforward
to interpolate the background in the range of the superconducting
transition with a smooth polynomial function. The total signal can
be represented as the complex variable ${\bf U} =
A_{S}e^{i\phi_{S}}$, and the interpolated background as ${\bf B} =
A{_B}e^{i\phi_B}$; our signal is then ${\bf
S}=A_{S}e^{i\phi_S}$={\bf U-B} (the difference of two complex
variables). The signal with background subtracted is shown in Fig.
\ref{fig9}b, which clearly separates into the main sharp peak
corresponding to the sample confined in the gasket hole with
T$_c\sim $16~K, and a broad peak at lower temperatures that is due
to the sample part which has flowed out of the gasket hole and is
confined between the diamond culet and the gasket.

\section{Simple Metals}
  We will focus in this section on simple $s-p$ metals. Theoretically these
metals  are considered well understood. The first treatment of simple
metals using pseudopotentials was given by McMillan
\cite{mcmil68}. This treatment was extended later by Allen and
Cohen \cite{AllenCohen}. By definition, ``simple metals''
are metals in which outer ($s$ and $p$) conduction electrons
are removed far enough in energy from $d$ or $f$ levels that these
conduction electrons can be treated as nonlocalized nearly free
electrons. In these metals the electron-ion interaction can be treated
using a pseudopotential approach, and the
pseudowave functions are similar to the free-electron plane waves. In
the following discussion we will use   the Allen-Dynes \cite{aldy}
modified McMillan's \cite{mcmil68} expression for T$_c$ :

\begin{equation} T_c={{\omega_{log} }\over{1.2}}{\rm
exp}\left({{-1.04(1+\lambda)}\over{\lambda-\mu^*(1+0.62\lambda)}}\right).
\label{eq:mcmil}
\end{equation}

\noindent Here the electron-phonon coupling constant is given by
\begin{equation}
\lambda=2\int {{d\omega \alpha^2 (\omega)F(\omega )}\over{\omega}}=
{{N(\varepsilon_F)<I^2>}\over{M<\omega^2>}},
\label{eq:lambda}
\end{equation}

\noindent where $\alpha(\omega)$ is an average of the  electron-phonon
interaction, $F(\omega)$ is the phonon density of states,
$N(\varepsilon_F)$ is the density of electron states at the Fermi
level, $<I^2>$ is the square of the electron-phonon interaction
matrix element averaged over the Fermi surface, $M$  is the atomic
mass, $\mu^\star$ is Morel-Anderson effective Coulomb repulsion
pseudopotential \cite{Morel}, and $\omega_{log}$ and $<\omega^2>$
are averages over the phonon spectrum  given by \cite{aldy}

\begin{eqnarray}
\omega_{log}=exp\left({{2}\over{\lambda}}\int_0^\infty {{dw}\over
{\omega}}\alpha^2F(\omega)ln\omega\right),\\
<\omega^2>={{2}\over{\lambda}}\int_0^\infty dw \alpha^2F(\omega)\omega.
\label{eq:wlog}
\end{eqnarray}

\noindent These expressions follow from a thorough analysis of the
dependence of the superconducting
transition temperature on material properties
($\lambda$, $\mu^\star$,  phonon spectrum)
as contained in Eliashberg theory \cite{mcmil68,aldy}.

Within the pseudopotential model, McMillan \cite{mcmil68} has derived 
the expression

\begin{equation}
<I^2>={{8}\over{9}}k_F^2E_F^2<v_q^2>,
\label{eq:Ife}
\end{equation}

\noindent where $E_F$ and $k_F$ are the Fermi energy and wavenumber,
and $<v_q^2>$ is a dimensionless average of the pseudopotential $V(q)$ squared

\begin{equation}
<v_q^2>=\int_0^{k_F} V(q)^2q^3dq / \int_0^{k_F} V(0)^2q^3dq
\label{eq:vq}
\end{equation}

\noindent For a free-electron gas, the density of states of one spin per atom
is $N(0)=3Z/4E_F$  \cite{mcmil68} ($Z$ is the valence of the atom). Expressing
the average phonon frequency in units of the ionic plasma frequency

\begin{equation}
\Omega_p^2=4\pi NZ^2e^2/M,
\label{eq:pl}
\end{equation}

\noindent McMillan \cite{mcmil68} gives the expression for the coupling constant as

\begin{equation}
\lambda={{N(0)<I^2>}\over{M<\omega^2>}}={{1}\over{2}}\pi
{{E_F}\over{k_Fe^2}} {{<v_q^2>}\over{(<\omega^2>/\Omega_p^2)}}.
\label{eq:lamfe}
\end{equation}

\noindent The factor $E_F/k_Fe^2$ is just 0.96/$r_s$, where $r_s$ is the radius in atomic units
of a sphere containing one electron; thus, a simple expression
can be derived \cite{mcmil68}

\begin{equation}
\lambda={{1.51}\over{r_s}} {{<v_q^2>}\over{(<\omega^2>/\Omega_p^2)}}.
\label{eq:lamsimp}
\end{equation}

\noindent McMillan \cite{mcmil68} noted that the  observed phonon
frequencies are extremely sensitive to small changes in
pseudoptential and the important dependence of the coupling constant
$\lambda$ upon the pseudopotential arises from the $<\omega^2>$
term in the denominator of Eq.\ref{eq:lamsimp}, rather than from
the $<v_q^2>$ in the numerator. Thus, for simple metals the
pseudopotential theory predicts that the coupling constant varies
inversely with the phonon frequency squared $\lambda\cong
C/(<\omega^2>/\Omega_p^2)$. Making further approximations,
McMillan arrives at his famous relation for simple metals:

\begin{equation}
\lambda\approx C'/M<\omega^2>.
\label{eq:lamC}
\end{equation}

Allen and Cohen \cite{AllenCohen} pointed out that in the jelium model

\begin{equation}
<\omega^2>/\Omega_p^2\approx {{1}\over{2}}q_D/k_s=(1/8r_s)(3\pi^2/Z)^{2/3}.
\label{eq:jelly}
\end{equation}

\noindent The potential strength $<v_q^2>$ is a  simple function
of $r_s$ in the jellium model, approximated by 0.075$r_s$ in the
region 2$<r_s<$5 (Fig. 4 in Ref. \cite{AllenCohen}). The net result
of the jellium model analysis is that $\lambda$ should scale as
$r_sZ^{2/3}$ on theoretical grounds, or as  $r_sZ^{5/3}$
empirically. We will use these results in our examination 
of the T$_c$ trends in chalcogen family given below. For the
analysis of high pressure data it is convenient to write down
an explicit expression for the Hopfield parameter $h=N(0)<I^2>$ as
follows from pseudopotential relation for $<I^2>$ from 
Eq.{\ref{eq:Ife}}

\begin{equation}
h=N(0)<I^2>={{3Z}\over{4E_F}}<I^2>={{2}\over{3}}Zk_F^2E_F<v_q^2>.
\label{eq:hfe}
\end{equation}

\noindent Using $<v_q^2>\sim r_s$, and $k_F^2E_F\sim 1/r_s^4$, we
find that $h\sim 1/r_s^3 \sim 1/V$, which is widely used by
experimentalists as an empirical relation for the Hopfield parameter in
simple metals. In the next sections we will explore how these
simple principles can be applied to real materials under pressure.

It should be noted, however,  that real metals deviate from this
simple model in many respects. As Allen and Cohen
\cite{AllenCohen} have pointed out, there are considerable
complications due to the following factors in real materials:
(i) The  actual Fermi surface is distorted away from a sphere near
zone boundaries.
(ii) The  actual matrix elements deviate from the free-electron
matrix elements near zone boundaries
(iii) The phonon frequencies are anisotropic.
We may add to this list:
(iv) The effects of anharmonicity play substantial role in
modifying electron-phonon  interaction (MgB$_2$, BaBiO$_3$, High-T$_c$
superconductors)
(v)  Different electron  energy bands (surfaces) have different
contributions to electron-phonon coupling in materials with
anisotropic energy bands (MgB$_2$).

\begin{figure}
\includegraphics[width=6in]{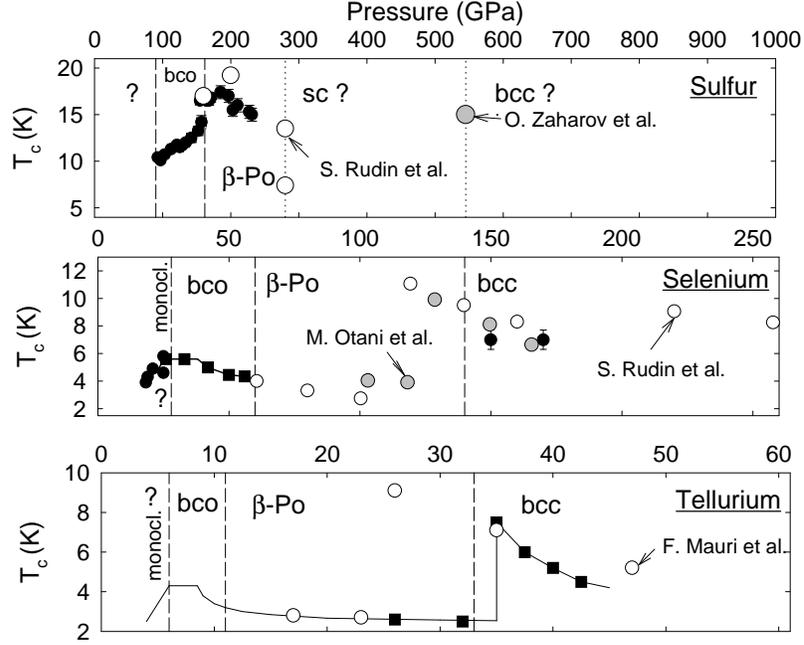} \caption{Superconducting T$_c$ in S, Se, and Te.
Data for S and Se below 25 GPa, and above 150 GPa are from
Refs. \cite{Sulfur97,Gregoryanz01}. Data for Te and the remaining
data for  Se are from Refs. \cite{Berman73,Bundy80,Akahama93}.
The observed similarity in T$_c$(P) curves illustrates the scaling
relations derived in the text. Theoretical calculations are also
shown (see Table \ref{tbl2}).} \label{fig10}
\end{figure}

\section{Chalcogens: Sulfur, Selenium, Tellurium}

We begin this section with estimates of scaling relations for
pressure as follows from chemical considerations. From the virial
theorem we have estimates for the total energy of valence
electrons in the form  \cite{Pauling}

\begin{equation}
W=\overline{T}+\overline{V}= -\frac{Z^2e^2}{2a_0 n^2}
\label{eq:virial}
\end{equation}

\noindent Here $\overline{T}$ is averaged kinetic energy, and
$\overline{V}$ is averaged potential energy, Z is the number of
valence electrons, $e$ is electron charge, $a_0$ is Bohr radius,
and $n$ is principal quantum number for the valence electrons ($n$=3
for S, $n$=4 for Se, $n$=5 for Te). According to Pauling \cite{Pauling},

\begin{equation}
W= -\frac{e^2a_0}{2n^2r^2},
\label{eq:coulomb}
\end{equation}

\noindent where $r\sim a_0$ is average electronic radius of the
valence orbitals. On the other hand, in the free-electron gas
approximation, the total energy of the free-electron gas is given
by the expression

\begin{equation}
E_F=(3\pi^2)^{2/3}(\hbar^2/2m)(Z/V_a)^{2/3},
\label{eq:Fermigas}
\end{equation}

\noindent where $V_a$ is atomic volume, (4/3)Z$\pi r_s^3$=$V_a$,
and $r_s$ is the radius of the sphere enclosing one electron.
Assuming that in metallic state the total energy $W\sim E_F$, we
find that  $n^2r^2 \sim V_a^{2/3}$, and thus $W\sim
1/(n^2V^{2/3})$, where V is the total volume. By noting that
$P\sim \frac{\delta W}{\delta V}$ we obtain $P\sim
1/(n^2r^5)$, ensuring the relation

\begin{table}
\caption{Scaling relations for chalcogen family. Atomic volumes at
phase transitions bcc $\mapsto$ $\beta$-Po  ($V_{\beta-Po}$) are
from Refs. \cite{Akahama93,Luo93}, atomic volumes at phase
transitions $\beta$-Po $\mapsto$ bcc  ($V_{bcc}$) are from Refs.
\cite{Akahama93,Zakharov}, pressure of the $\beta$-Po $\mapsto$
bcc phase transition for sulfur is taken from theory
\cite{Zakharov} }
\begin{tabular} {cccc}
Element (n) & Te (5)& Se (4) & S (3)\\
\hline
$V_{\beta -Po}$ (\AA$^3$) & 25.8&14.3 & 8.5 \\
$r_v=r_sZ^{1/3}$  (\AA)& 1.83&1.51&1.27\\
$P$ (GPa) & 10.5 & 60 & 160 \\
$n^2r_v^5P$ (\AA$^5\cdot$GPa) & 5.4$\cdot$10$^3$ &
7.5$\cdot$10$^3$ &
4.7$\cdot$10$^3$ \\
\hline
$V_{bcc}$ (\AA$^3$) & 21 &11.3& 7.4 \\
$r_sZ^{1/3}$  (\AA)& 1.71&1.39&1.21\\
$P$ (GPa) & 33 & 140 & 545 \\
$n^2r_v^5P$ (\AA$^5\cdot$GPa) & 1.2$\cdot$10$^4$ &
1.2$\cdot$10$^4$ &
1.3$\cdot$10$^4$ \\
\hline
\end{tabular}
\label{tbl1}
\end{table}

\begin{table}
\caption{Average phonon frequencies and electron-phonon coupling
in the chalcogen family superconductors at selected pressures from
theoretical work \cite{Zakharov,rudin,Mauri,Otani}}
\begin{tabular} {cccccc}
P (GPa)& $\omega_2$ (cm$^{-1}$)&    $\omega_{log}$(cm$^{-1}$) &$\lambda$ & T$_c$(K)&  Ref.\\
\hline
& & Sulfur & & & \\
160 ($\beta$-Po)&  375 & 305 & 0.76 &   17 & \cite{rudin} \\
280 ($\beta$-Po) &463& 343 & 0.66& 13.5   & \cite{rudin} \\
280 (sc) & 481 & 389 & 0.53& 7.4 & \cite{rudin} \\
584 (bcc) & 411 &422 &0.58 &15& \cite{Zakharov}\\
\hline
 &   &      Selenium &  &  & \\
103 ($\beta$-Po) & 217 & 174&  0.58  &  4.04  &  \cite{Otani} \\
118(($\beta$-Po)&  227 & 179 & 0.57 & 3.91 & \cite{Otani} \\
129 (bcc) & 204 & 157 & 0.83&  9.9 & \cite{Otani}\\
166 (bcc) & 234 & 185 & 0.66 & 6.6 &\cite{Otani}\\
\hline
& &  Tellurium & & & \\
17 ($\beta$-Po) &   109 & 57 & 0.8 & 2.8 & \cite{Mauri}\\
23 ($\beta$-Po)  & 116 & 55 & 0.8 & 2.7 & \cite{Mauri} \\
26 (bcc) &   103&  87 & 1.64 & 9.1 & \cite{Mauri} \\
47 (bcc) & 131 & 58 & 0.93 & 5.2 & \cite{Mauri}\\
 \hline
\end{tabular}
\label{tbl2}
\end{table}

\begin{equation}
n^2r^5P=const
\label{eq:Pscaling}
\end{equation}

\noindent holds through the chalcogen series for equivalent
free-electron-like metallic states as long as there is no
substantial contribution from s- or d-orbitals to the energy
balance. We have estimated that the product in Eq.
\ref{eq:Pscaling} is indeed almost invariant for phase transition
pressures in the chalcogen series from bco to $\beta$-Po and from
$\beta$-Po to bcc in metallic phases (illustrated by data
in Table \ref{tbl1}). We also show the behavior of the
superconducting  T$_c$ for S, Se, and Te in
Fig. \ref{fig10} by adjusting the pressure scale to the bcc phase
transition pressure point. The observed similarities in T$_c$(P)
mimic similarities in the phase transition sequence.
Note also that theoretically predicted T$_c$'s for the $\beta$-Po and
bcc phases are in good agreement with experiment. Table
\ref{tbl2} shows theoretical estimates for averaged phonon
frequencies and electron-phonon coupling $\lambda$ as follows from
theoretical calculations. It is remarkable  that the $\lambda$ values change
very little when going from Te to S (except for the very high $\lambda$
in Te bcc phase close to the phase transition, which is attributed
to soft phonon modes  \cite{Mauri}). Thus, instead of
$\lambda\sim r_s$ as we would expect from pseudopotential theory
(see Ref. \cite{AllenCohen}, and discussion after Eq. \ref{eq:jelly} in
section 3), we have instead $\lambda\sim$const. If this trend
continues to oxygen, we may expect rather high T$_c$ values in
atomic oxygen at pressures above 1000 GPa \cite{OtaniO}.

\section{MgB$_2$ and Phonon-Assisted Electronic Topological Transition}

The recently discovered high-temperature superconductor MgB$_2$
\cite{akimitsu} has attracted considerable interest. 
Experiment \cite{bud'ko,hinks}
as well as theory \cite{kortus,an,kong,yildirim} indicates that
MgB$_2$ can be treated as a phonon mediated superconductor.
Calculations show that the strongest coupling is realized for the
phonon branch in the Brillouin zone from $\Gamma$ ($E_{2g}$
phonon) to $A$ ($E_{2u}$ phonon), which is  related to
vibrations of the B atoms \cite{an,kong,yildirim}. This makes
MgB$_2$ a unique system in which a single phonon branch appears to dominate the
superconducting properties within the framework of a phonon-mediated
mechanism for superconductivity. By knowing the pressure
dependence of these phonon frequencies and the pressure dependence of
$T_c$, the electron-phonon coupling in this material can be
directly addressed.

\begin{figure}
\includegraphics[width=4in]{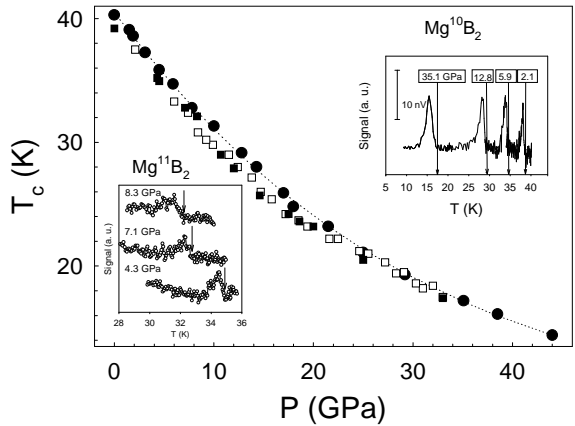} \caption{This figure shows
pressure dependence of T$_c$ in isotopically pure MgB$_2$ samples,
measured in He pressure medium. Corresponding insets show
experimental temperature scans. Circles are used for Mg$^{10}$B$_2$,
squares for Mg$^{11}$B$_2$; full symbols are for compression and  open 
symbols for decompression.  } \label{fig11}
\end{figure}

\begin{figure}
\includegraphics[width=5in]{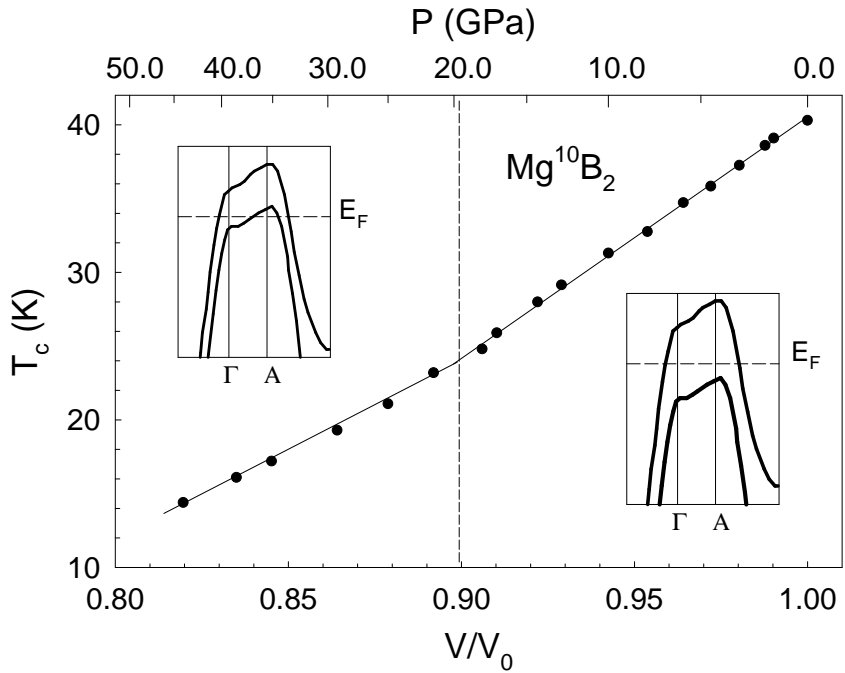} \caption{Superconducting 
critical temperature ($T_c$) plotted
as a function of volume is linear, with kink at about 20
GPa for Mg$^{10}$B$_2$, respectively. Insets illustrate the proposed
phonon-assisted electronic topological transition, responsible for 
the observed effect. See text for details. }
\label{fig12}
\end{figure}

We measured T$_c$(P) in isotopically pure samples of MgB$_2$
(samples were  similar to those used in Ref.\cite{bud'ko}), with
$^{10}$B and $^{11}$B. The details of the experiment are presented
in Ref. \cite{strumgb2}. In Fig.\ref{fig11} we show the $T_c$ as a
function of pressure; temperature scans at selected pressures are
also shown. The signal observed is close to the limit of the
sensitivity of our setup. The signal is superimposed on the
nonlinear paramagnetic background from the gasket material at
lower temperatures (below 25 K), which has a characteristic
${{1}\over{T}}$ dependence (subtracted from the data for
Mg$^{10}$B$_2$. However, the onset of $T_c$ can be reliably
identified with an accuracy  0.2-0.8 K (depending on the actual
quality of the data, as illustrated in Fig.\ref{fig11}) up to the
highest pressures reached in this experiment.

We plot $T_c$ for Mg$^{10}$B$_2$ as a function
of volume in Fig.\ref{fig12}. One can clearly distinguish a kink
in $T_c$(V) at a volume that corresponds to 20 GPa. We observe 
a similar kink at 15 GPa for Mg$^{11}$B$_2$  \cite{strumgb2}. 
 Pressure dependence of the $E_{2g}$ phonon was
measured recently in our laboratory \cite{gonchmgb2}. 
We have measured $E_{2g}$ Raman mode frequency
in Mg$^{10}$B$_2$ at room temperature to 50 GPa \cite{strumgb2} to
understand the anomaly in $T_c$; we observed similar anomaly in
the pressure dependence of the Raman mode slightly above 20 GPa. The
details of the Raman experiment are published elsewhere
\cite{alexraman}.

At lower
pressures the zero-point motion of the boron atoms for the $E_{2g}$ mode
strongly splits the boron in-plane $\sigma$ bands, so that the
lower band moves below the Fermi level, thereby crossing it and
fulfilling the condition for an Electronic Topological Transition (ETT) 
(Fig. \ref{fig12}, right inset). 
This means that for a
frozen-phonon calculation there should be an anomalous
contribution to the  total energy, that behaves similarly to
 $2{{1}\over{2}}$ power  term in the free energy, suggested by 
Lifshitz \cite{Lifshitz}, with the amplitude of the phonon mode being a
parameter that drives the electronic subsystem through the
transition. As such, the phonon frequency is strongly
anharmonic, and its volume derivative (the Gr\"uneisen parameter)
may even diverge at the transition. At higher pressures
the zero-point motion does not split  the $\sigma$ band strongly enough
for the lower band to cross the Fermi level (Fig. \ref{fig12}, left inset). 
Thus, the system is always at conditions in which there is no anomalous
contribution to the free energy from ETT-like  2${{1}\over{2}}$
power terms, and the phonon mode and $T_c$ behave in a more
regular manner. Between those two regimes there should be a small
pressure range in which the amplitude of the zero-point motion is
just enough for the top of the lower band to coincide with the
Fermi level. It is proposed that this condition is almost fullfiled at
the observed kinks in pressure dependencies of $T_c$ and $E_{2g}$
phonon frequency.

 The lower pressure for the observed transition
 in Mg$^{11}$B$_2$ versus Mg$^{10}$B$_2$  \cite{strumgb2} 
may be due to the isotope effect: the zero-point motion for a heavier
atom is smaller, and the matching condition for the  $\sigma$ band
is fulfilled at lower compression of the lattice. Several arguments 
support the observed isotope trend, following
the reasoning proposed by An and Pickett \cite{an}. They noticed
that the $\sigma$  bands belonging to B, which form
cylindrical Fermi surface sheets  \cite{kortus}, can be treated as
quasi-two-dimensional. The states in these bands contribute most
of the electron-phonon coupling responsible for superconductivity
\cite{an}. The overall splitting of the $\sigma$ band is
characterized by $p-p$ matrix element $t_{pp\sigma} \sim d^{-3}$,
where $d$ is B-B bond length. Thus, the deformation potential of
the $\sigma$ band  will be proportional to the derivative of the
above matrix element with respect to $d$ ($E_{2g}$ mode modulates
B-B distance), and thus  $|\vec {\cal D}| \sim d^{-4}$. We have
determined earlier that the phonon frequency of the E$_{2g}$ mode
scales as $\omega \sim (a/a_0)^{-10.8}= (d/d_0)^{-10.8}$ below 15
GPa  \cite{gonchmgb2}. The amplitude of the zero-point motion $u
\sim \omega^{-1/2}\sim (d/d_0)^{5.4}$, which means that the
splitting of  $\sigma$  bands $\Delta E \sim \vec {\cal D} u \sim
(d/d_0)^{1.4}$ decreases almost proportionally with the B-B bond
length.

The low-pressure regime for both isotopic compounds suggests a
strong contribution from ETT anomalies to the observed properties
of the materials. It should be noted that the electron-phonon
coupling may be strongly affected by the non-adiabatic effects due
to the violation of the condition that the Debye frequency is much
less than the Fermi energy  $\omega_D/E_F \ll 1$ \cite{Migdal}
close to the ETT regime. We also expect large effects of uniaxial
stresses and stoichiometry  on the pressure dependence of
the superconducting transition in MgB$_2$.

\section{Conclusions}

Developments in magnetic susceptibility techniques for  diamond
anvil cell applications have beeen reviewed here with an in-depth
discussion of the basics of the recently improved double modulation 
technique. The method  has been applied to studies above
230~GPa without a pressure medium and above 40~GPa with He
pressure medium. The pressure-limiting factors have not been yet
explored in detail, and high-pressure limit of the technique
remains to be established. Numerical estimates/extrapolations
indicate that superconductivity measurements on samples down to
10~m$\mu$ in diameter are within reach \cite{Timofeev00}.

We have measured superconducivity in S over broad pressure
range from metalization onset at 90~GPa (bco phase) to 231~GPa
into the $\beta$-Po phase. Superconductivity in Se was studied
below 33~GPa, and the pressure-induced increase of T$_c$ was
discovered in a yet to be identified low-pressure phase; we also
observed T$_c$ $\sim$7.5~K above 160~GPa. These
experiments, together with previous pressure studies of
superconductivity in Te and Se, indicate  that pressure-induced
changes of T$_c$ in the chalcogens follow a scaling relation 
$n^2r_s^5P\approx$const. (Eq. \ref{eq:Pscaling}).
The electron-phonon coupling constant in the chalcogens in the
corresponding phases appear to be remarkably similar, a result that  
can not be explained by the trends derived from simple 
pseudopotential model.

MgB$_2$ is an example of a less isotropic simple s-p metal, with
a T$_c$ value almost at the limit of what is expected from
a phonon-mediated superconductor. We have explored the pressure
dependence of T$_c$ for two B isotopes: Mg$^{10}$B$_2$ and
Mg$^{11}$B$_2$. Combining T$_c$ studies with Raman measurements of
the E$_{2g}$ mode, we were able to find a correlation between the
highly anharmonic behavior of the phonon  and T$_c$(P) anomaly
in Mg$^{11}$B$_2$. A Lifshitz ETT \cite{Lifshitz}, with the amplitude of the
zero point motion of the B atoms taken into account \cite{strumgb2}, 
can explain the origin of these effects.

\acknowledgments
This work was supported by the National Science
Foundation, the Department of Energy, the Center for High Pressure
Research (CHiPR). Special thanks are to the former Director of the 
Geophysical Labotatory Charles Prewitt, who endorsed and supported 
this work under the auspices of CHiPR.

\end{document}